\documentclass[10pt,a4paper]{article}
\usepackage{a4wide}
\usepackage{graphicx}
\usepackage{hyperref}
\usepackage{url}
\usepackage{color}

\begin{document}
\title{Cloud for Gaming\footnotemark}
\author{Gabriele D'Angelo \qquad Stefano Ferretti \qquad Moreno Marzolla\\
University of Bologna, Dept. of Computer Science and Engineering\\
Mura Anteo Zamboni 7, I-40126 Bologna, Italy\\
\texttt{\{g.dangelo, s.ferretti, moreno.marzolla\}@unibo.it}}
\date{}

\maketitle

\footnotetext{The publisher version of this book chapter is available at \url{http://dx.doi.org/10.1007/978-3-319-08234-9_39-1}.
\textbf{{\color{red}Please cite this book chapter as: ``Gabriele D'Angelo, Stefano Ferretti, Moreno Marzolla. Cloud for Gaming. Encyclopedia of Computer Graphics and Games. Springer International Publishing, 2015, ISBN 978-3-319-08234-9.''}}}

\section*{Definition}

Cloud for Gaming refers to the use of cloud computing technologies to
build large-scale gaming infrastructures, with the goal of improving
scalability and responsiveness, improve the user's experience and
enable new business models.

\section*{What is cloud computing?}

Cloud computing is a service model where the provider offers
computation and storage resources to customers on a ``pay as you go''
basis~\cite{mell11}. The essential features of a cloud computing
environment are:

\begin{description}
\item[On-demand self service:] the ability to provide computing
  capabilities (e.g., CPU time, network storage) dynamically, as
  needed, without human intervention;

\item[Broad network access:] resources can be accessed through the
  network by client platforms using standard mechanisms and protocols;

\item[Resource pooling:] virtual and physical resources can be pooled
  and assigned dynamically to consumers, according to their demand,
  using a multi-tenant model;

\item[Elasticity:] from the customers point of view, the provider
  offers unlimited resources that can be purchased in any quantity at
  any time;

\item[Measured service:] cloud resource and service usages are
  optimized through a pay-per-use business model, and are monitored,
  controlled and reported transparently to both their customer and
  provider.
\end{description}

The typical interaction between cloud provider and customer works as
follows: the customer connects to a ``cloud marketplace'' through a
Web interface, and selects the type and amount of the resources she
needs (e.g., some virtual servers with given number of CPU cores,
memory and disk space). The resources are allocated from a large pool
that is physically hosted on some big datacenter managed by the cloud
provider. Once instantiated, the resources are accessed by the
customer through the network. Additional resources can be acquired at
a later time, e.g., to cope with an increase of the workload, and
released when no longer needed. The customer pays a price that depends
on the type and amount of resources requested (e.g., CPU cores speed,
memory size, disk space), and on the duration of their usage.

\begin{figure}[t]
  \centering%
  \includegraphics[width=.6\textwidth]{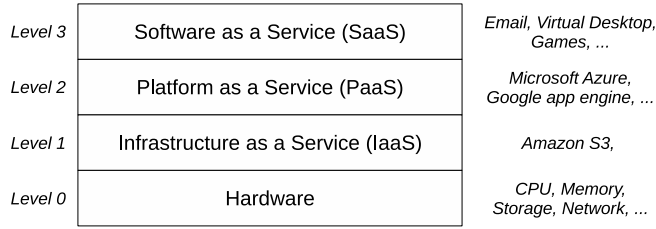}
  \caption{Cloud Service Model}\label{fig:service-model}
\end{figure}

The \emph{service model} defines the level of abstraction at which the
cloud infrastructure provides service
(Figure~\ref{fig:service-model}). In a \emph{Software as a
  Service}~(SaaS) cloud, the system provides application services
running in the cloud. ``Google apps'' is an example of a widely used
SaaS cloud.  In contrast, the capabilities provided by a
\emph{Platform as a Service}~(PaaS) cloud consist of programming
languages, tools and a hosting environment for applications developed
by the customer. The difference between the~SaaS and~PaaS models is
that while the user of a~SaaS cloud simply utilizes an application
that runs in the cloud, the user of a~PaaS cloud develops an
application that can be executed in the cloud and made available to
service customers; the application development is carried out using
libraries, APIs and tools possibly offered by some other
company. Examples of~PaaS solutions are AppEngine by Google, Force.com
from SalesForce, Microsoft's Azure and Amazon's Elastic
Beanstalk. Finally, an \emph{Infrastructure as a Service}~(IaaS) cloud
provides its customers with fundamental computing capabilities such as
processing, storage and networks where the customer can run arbitrary
software, including operating systems and applications. The number of
companies offering such kind of services is continually growing; one
of the earliest being Amazon with its~EC2 platform.

The \emph{deployment model} defines the mode of operation of a cloud
infrastructure; these are the private cloud, the community cloud, the
public cloud, and the hybrid cloud models. A \emph{private cloud} is
operated exclusively for a customer organization; it is not
necessarily managed by that organization. In the \emph{community
  cloud} model the infrastructure is shared by several organizations
and supports a specific community with common concerns (e.g., security
requirements, policy enforcement). In the \emph{public cloud} model
the infrastructure is made available to the general public and is
owned by an organization selling cloud services. Finally, the
\emph{hybrid cloud} model refers to cloud infrastructures constructed
out of two or more private, public or community clouds. 

\section*{Cloud computing for gaming}

The gaming industry embraced the cloud computing paradigm by
implementing the \emph{Gaming as a Service}~(GaaS) model~\cite{Cai14}.
Different instances of the GaaS paradigm have been proposed: remote
rendering~GaaS, local rendering~GaaS and cognitive resource
allocation~GaaS.

\begin{figure}[t]
\centering\includegraphics[width=.8\textwidth]{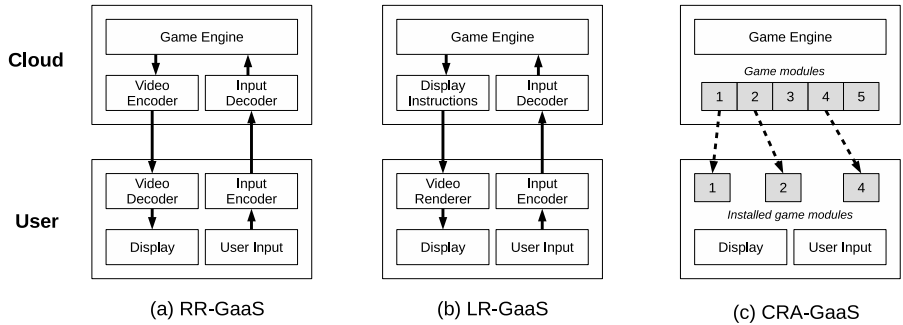}
\caption{Gaming as a Service models}\label{fig:gaas-models}
\end{figure}

In the \emph{remote rendering}~GaaS (RR-GaaS) model the cloud
infrastructure hosts one instance of the game engine for each player
(Fig.~\ref{fig:gaas-models}~(a)). An encoder module running on the
cloud is responsible for rendering every frame of the game scene, and
compressing the video stream so that it can be transmitted to the
user's terminal where the stream is decoded and displayed. User inputs
are acquired from the terminal and sent back to the game engine that
takes care of updating the game state accordingly. The advantage of
the~RR-GaaS model is that the workload on the terminal is greatly
reduced, since the computationally demanding step of rendering the
game scenes is entirely offloaded to the cloud. This allows complex
games to be played on less powerful devices, such as mobile phones or
cheap game consoles, that are only required to be capable of decoding
the video stream in real time. However, the RR-GaaS model consumes
considerable bandwidth to transmit the compressed video stream, and
may be particularly sensitive to network delays. Examples of RR-GaaS
implementations are GamingAnywhere~\cite{Huang14} and Nvidia
Grid\textsuperscript{TM}\footnote{\url{http://www.nvidia.com/object/cloud-gaming.html},
  Accessed on 2015/4/4}.

In the \emph{local rendering} GaaS model, the video stream is encoded
on the cloud as a sequence of high-level rendering instructions that
are streamed to the player terminal (Fig.~\ref{fig:gaas-models}~(b));
the terminal decodes and executes the instructions to draw each
frame. Since encoding of each frame as a sequence of drawing
instructions is often more space efficient than compressing the
resulting bitmap, the LR-GaaS model may require less network bandwidth
than RR-GaaS, and therefore eliminate the need for real-time video
transmission capability. This comes at the cost of requiring a more
powerful terminal with an adequate graphics subsystem.

Finally, in the \emph{cognitive resource allocation}~GaaS model, the
game engine is logically partitioned into a set of modules that can be
upload and executed at the client side
(Fig.~\ref{fig:gaas-models}~(c)). As the game evolves, the terminal
receives and executes the appropriate modules, and may keep or discard
the unused ones. The CRA-GaaS model shifts the computation back to the
client terminal, therefore reducing the load on the cloud. However,
the client resources are used efficiently, since at any time only the
needed components are stored locally. This is a significant advantage
if we consider that the data of a complete modern game takes a lot of
space for textures, 3D models, sounds and code modules.

GaaS provides advantages for both game developers and players. The
ability to offload some computation on the cloud allows simple
terminals such as mobile devices to play complex games. Since the
game engine is accessed on demand, flexible business models such as
pay-per-play or monthly subscription can be easily
implemented. Finally, game operators can scale up and down the amount
of cloud resources used by the gaming infrastructure.

\begin{figure}[t]
  \centering%
  \includegraphics[width=.8\textwidth]{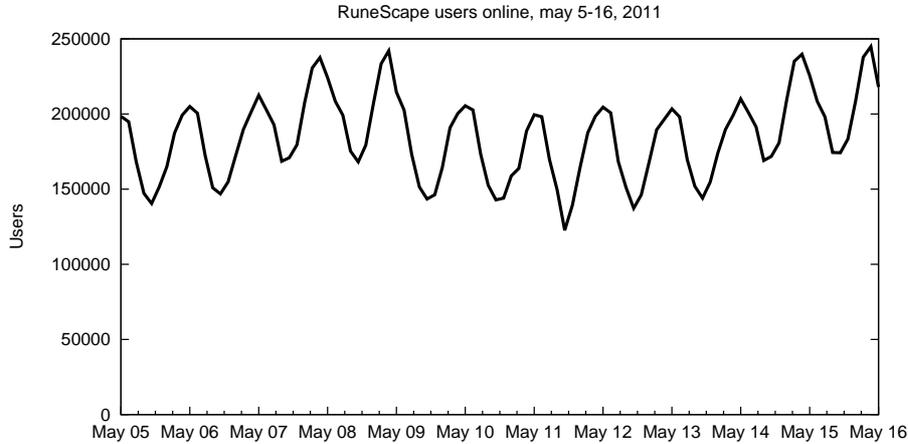}
  \caption{Number of online players of the Runescape~MMOG; the data
    refers to the period from may~5 to may~16,
    2011}\label{fig:runescape-users}
\end{figure}

The last point is particularly important, especially for the so-called
\emph{Massively Multiplayer Online Games} (MMOG). Modern~MMOGs are
large-scale distributed systems serving millions of concurrent users
which interact in real-time with a large, dynamic virtual world.  The
number of users playing the game at any given time follows a pattern
that originates from the typical daily human activity. As an example,
Figure~\ref{fig:runescape-users} shows the number of online players of
RuneScape\footnote{\url{http://www.runescape.com}}~\cite{Marzolla12},
a fantasy game where players can travel across a fictional medieval
realm. During the observed period, more than 200,000 players are
connected to the system at peak hours; this number reduces to about
110,000 players during off-peak hours. Hence, the daily churn (number
of players leaving/joining the system during the day) is about 100,000
users. It is evident that static resource provisioning based on the
average load results in system overloaded roughly half the time;
provisioning for the worst case results in a massive resource
under-utilization.

To effectively implement a cloud-based gaming infrastructure, it is
necessary to address non-trivial issues related to game state
partitioning, responsiveness, synchronization and security.

\paragraph{Partitioning}
The key factor for achieving scalability of a~GaaS infrastructure is
the ability to partition the workload across the cloud resources. This
is relatively easy if the workload consists of the execution of
independent game instances that can be executed on any available
resource, irrespective of where other instances are running. This is
the case when the game does not allow different players to
interact. Things become complex if the instances are \emph{not}
independent, as in the case of a~MMOG system where all players
interact with the same virtual world. In this case, the game engine
must maintain a large shared state, allowing the players to ``see''
the effects of actions performed by the other players operating in the
same virtual location.

This is achieved by partitioning the virtual world across multiple
zones, each handled by a separate set of cloud resources. Given that
communication between resource instances may incur significant delays,
it is important that interaction across neighboring zones is
minimized. For example, each partition may hold a collection of
``islands'' such that all interactions happen within the collection,
while players can jump from one ``island'' to another. 

Depending on the (virtual) mobility pattern of each player, some areas
of the game field may become crowded, while others may become less
populated. In order to cope with this variability, each zone
controller is physically hosted on resources provided and operated by
a cloud infrastructure. The cloud provider is in general a separate
entity providing computational and storage resources to the game
operator on a pay-as-you go model. This means that the game operator
can request additional servers and/or additional storage space at any
time, and release them when no longer needed. Thus, the game operator
can request more resources when the workload on a zone increases, in
order to keep the response time perceived by players below a
predefined maximum threshold. When the workload decreases, the game
operator can release surplus resources in order to reduce costs.

\paragraph{Synchronization}
The success of a gaming system is based on having players perceiving
the game state as identical and simultaneously evolving on every
player participating to a gaming session. If the game state is
replicated in different cloud servers, a synchronization algorithm is
needed to maintain the consistency of the redundant game state. To
this aim, different schemes have been proposed in the
literature~\cite{Furht06}. They mainly differ from classic
synchronization algorithms employed by distributed systems in their
additional requirement for keeping the computation quick and
responsive. To this aim, some schemes relax the requirements for full
consistency during the game state computation.

A basic distinction is between conservative and optimistic
synchronization. Conservative synchronization approaches allow the
processing of game updates only when it is consistency-safe to do
so. Lockstep~\cite{Fujimoto:1999}, time-bucket
synchronization~\cite{Fujimoto:1999}, interactivity
restoring~\cite{Ferretti:chap}, are some examples in the literature.

Optimistic synchronization mechanisms process game updates as soon as
they receive them, thus increasing the responsiveness of the
system. Yet, it is assumed that most updates are received in the
correct order and that, in any case, it would be acceptable to recover
later from possible inconsistencies.  Examples of optimistic
approaches available in the scientific literature are the optimistic
bucket synchronization~\cite{diot}, the combination of local lag and
Time Warp proposed in~\cite{Mauve:2002}, the trailing state
synchronization~\cite{Cronin:2002}, and the improved Time Warp
equipped with the dropping scheme and a correlation-based delivery
control approach~\cite{Ferretti:chap}.

\paragraph{Responsiveness}
The task of providing a pleasant experience to players becomes
challenging when trying to deploy a large scale and highly interactive
online game. Responsiveness means having small delays between the
generation of a game update at a given player and the time at which
all other players perceive such update.  How much such delays must be
small depends on the type of online game.  Obviously, the shorter the
delay the better is. But it is possible to identify a a game-specific
responsiveness threshold $T_r$ that represents the maximum delay
allowable before providing a game update to players. The typical $T_r$
for fast-paced games (e.g., first-person shooter, racing vehicles) is
150 to 200ms, but this value can be increased to seconds in slow
paced-games (e.g., strategic, role-playing
games)~\cite{Ferretti:chap,Pantel:2002}.

A key point is that each player is geographically distributed. Thus,
his latency to reach the game server on the cloud is usually different
from other players. If a classic client-server approach is employed,
it might thus happen that a responsive service is provided to some
subset of users, while the other players can perceive a non responsive
game evolution.  This raises another main issue, i.e.~fairness
provision. This means guaranteeing that all players have the same
chance of winning, regardless of their subjective network
conditions~\cite{Ferretti:chap}. To this aim, it should be guaranteed
that all players perceive the same and simultaneous game evolution at
the same time.

GaaS infrastructures represent an effective tool to provide
responsive and fair gaming experiences. Cloud servers can manage the
game state evolution in a scalable manner. Multiple server instances
can be run in the same datacenter, when needed. Moreover, if the game
involves world wide distributed players, one might think to introduce
a federation of cloud servers, geographically distributed, so that
each client/player might connect to its nearest server. This could
balance the network delays between the player and its server, thus
augmenting the fairness level provided by the system.  However, when
multiple servers are involved, each one with a redundant copy of
the game state, synchronization algorithm are needed to maintain game 
state consistency.

\paragraph{Security and reliability}
The security issues of GaaS infrastructures have become mainstream
after the PlayStation Network outage that, in 2011, has halted the
Sony online gaming network for 23 days. The network was shut down
after detecting an external intrusion that led to a huge number of
accounts being compromised, and the exposure of the players' personal
information.

From the reliability point of view, large cloud systems provide some
level of redundancy to cope with failures, including the use of
geographically distributed datacenters, so that catastrophic events do
not cause a complete outage. Unfortunately, the GaaS infrastructure
may still represent a single point of failure; the PlayStation Network
outage is just one example: in that case a security incident prompted
the system administrators to temporarily shut down the whole service.
Other possibilities must be considered as well: for example, the
company operating the GaaS infrastructure may go bankrupt, depriving
all players from the game service they might already have paid for.

From the security point of view, GaaS infrastructures are affected by
the typical issues of cloud computing (e.g. insider
attacks~\cite{Zissis2012583}) and online gaming
(e.g. cheating~\cite{SEC:SEC5}). Online games are an appealing target
for hacks because players often invest huge amount of time in their
character development, and is therefore quite easy to monetize game
items on the black market. Additionally, individual accounts on online
gaming platforms often contains information, such as credit card
numbers, that are the typical target of cyber-criminals. Details of
the avatar of each player can provide information such as sexual
preferences~\cite{gender} that could cause considerable embarrassment
if made public.

\end{document}